\documentclass[referee]{raa}            

\usepackage{graphicx,times}             
\usepackage{natbib}
\usepackage{amssymb,amsmath}
\bibpunct{(}{)}{;}{a}{}{,}

\newcommand{\angstrom}{\,\textup{\AA}}
\usepackage[a4paper=true,pagebackref=true]{hyperref}
\hypersetup{colorlinks = true, linkcolor = green, anchorcolor = red, citecolor = blue, filecolor = red, pagecolor = red, urlcolor = red}

\begin{document}

   \title{Radiative Hydrodynamic Simulations of the Spectral Characteristics of Solar White-light Flares
}

   \volnopage{Vol.0 (20xx) No.0, 000--000}      
   \setcounter{page}{1}          

   \author{Y. T. Yang
      \inst{1,2}
   \and Jie Hong
      \inst{3}
   \and Y. Li
      \inst{1,2}
   \and M. D. Ding
      \inst{3}
   \and H. Li
      \inst{1,2}
   }

   \institute{Key Laboratory of Dark Matter and Space Astronomy, Purple Mountain Observatory, Chinese Academy of Sciences, Nanjing 210033, People's Republic of China; {\it yingli@pmo.ac.cn}\\
        \and
             School of Astronomy and Space Science, University of Science and Technology of China, Hefei 230026, People's Republic of China\\
        \and
             School of Astronomy and Space Science, Nanjing University, Nanjing 210023, People's Republic of China; {\it jiehong@nju.edu.cn}\\
\vs\no
   {\small Received~~2020 June 10; accepted~~2020~~June 26}}

\abstract{ 
As one of the most violent activities in the solar atmosphere, white-light flares (WLFs) are generally known for their enhanced white-light (or continuum) emission, which  primarily originates in the solar lower atmosphere. However, we know little about how white-light emission is produced. In this study, we aim to investigate the response of the continua at 3600\angstrom\ and 4250\angstrom\ and also the H$\alpha$ and Ly$\alpha$ lines during WLFs modeled with radiative hydrodynamics simulations. We take non-thermal electron beams as the energy source for the WLFs in two different initial atmospheres and vary their parameters. Our results show that the model with non-thermal electron beam heating can clearly show enhancements in the continua at 3600\angstrom\ and 4250\angstrom\ as well as in the H$\alpha$ and Ly$\alpha$ lines. 
A larger electron beam flux, a smaller spectral index, or a penumbral initial atmosphere leads to a stronger emission increase at 3600\angstrom, 4250\angstrom\ and in the H$\alpha$ line. For the Ly$\alpha$ line, however, it is more preferably enhanced in a quiet-Sun initial atmosphere with a larger spectral index of the electron beam. It is also notable that the continua at 3600\angstrom\ and 4250\angstrom\ and the H$\alpha$ line exhibit a dimming at the beginning of the heating and reach their peak emissions later than the peak time of the heating function, while the Ly$\alpha$ line does not show such behaviors. These results can be served as a reference for analyzing future WLF observations.
\keywords{methods: numerical --- radiative transfer --- Sun: atmosphere --- Sun: flares}
}


   \maketitle

%
%
\section{Introduction}           
\label{sect:intro}

It is generally acknowledged that solar flares are one of the most interesting and intriguing phenomena among the solar activities. And solar white-light flares (WLFs) are one type of solar flares characterized by an enhancement in the optical continuum. They are regarded to be the most violent and energetic flare events. However, in general, WLFs are relatively rare since the first report by \citet{carrington1859description}, with a very small recorded number compared to the total recorded solar flares. The first space observation of WLFs was reported by \citet{hudson1992white} using the Soft X-ray Telescope (Yohkoh/SXT; \citealt{tsuneta1991soft}). WLFs are more prone to arise in the neighborhood of sunspots which can facilitate the emission enhancement in visible continuum and originate in the solar lower atmosphere \citep{ding1996possible,hudson2006white,jess2008do,wang2009study}. A comprehensive review on the WLF research history has been presented by \citet{hudson2016chasing}.

With the emergence of more ground-based white-light telescopes and solar satellites which can observe in a waveband including the optical part, such as Yohkoh/SXT, the Transition Region and Coronal Explorer (TRACE) and Solar Optical Telescope (Hinode/SOT), more studies have been aimed at the spectral characteristics and possible mechanisms of WLFs. \citet{liu2001enhanced} found that a flare on  2001 March 10 with a continuum enhancement near the Ca {\footnotesize{II}}  8542\angstrom\ line exhibits a temporal correlation with the peak of the flux at 7.58\,GHz. \citet{chen2005relationship} showed that the continuum enhancement has a fine co-temporal feature with the hard X-ray (HXR) emission produced by bremsstrahlung \citep{brown1971deduction} of the non-thermal electrons. \citet{jing2008spatial} studied a WLF on 2006 December 13 and the results show that the continuum emission enhancements during WLFs are co-spacial to the place where the most energy is deposited from non-thermal electrons which are accelerated by magnetic reconnection \citep{neidig1993energetics, matthews2003catalogue, metcalf2003trace, hudson2006white,chen2006footpoint,fletcher2007trace}. Moreover, \citet{neidig1993solar}, \citet{xu2006high} and \citet{isobe2007flare} indicate that there is often a core-halo structure of WLF emission, namely a brighter core emission encircled by a fainter and diffused halo. This observational property implies complex components of WLF emission that are related to different heating mechanisms, i.e., direct heating by non-thermal electrons and back-warming. \citet{namekata2017statisticala} studied 50 solar WLFs and examined the correlation between their energies and durations. The result reveals that this relation is similar to that on stellar superflares, in spite of the longer durations but much less energies of solar WLFs than those of stellar superflares. \citet{hao2017circular} identified white-light emission in a circular flare and found an impulsive and a gradual white-light kernels \citep{kane1985characteristics} which might be related to different heating mechanisms. \citet{song2018observations} found a standard fan-spine magnetic field configuration \citep{sun2013hot} of a circular-ribbon flare with white-light emission.

The possible radiation mechanisms of white-light continuum include recombination radiation (free-bound radiation) of hydrogen atoms in the lower chromosphere and the negative hydrogen ion emission in the upper photosphere \citep{ding1994optical}. According to the different spectral features from observational data, there are two types of WLFs which are suspected to have different emission mechanisms (e.g., \citealt{machado1986mechanism}). Type I WLFs show a Balmer jump, strong emission of hydrogen lines, and a temporal correlation with HXR emission, while type II WLFs hardly have such features except that \citet{prochazka2018reproducing} stated that type II WLFs could also show a temporal correlation between the Balmer and the HXR emission.

Type I WLFs may be heated by non-thermal electron beams followed by radiative back-warming. Type II WLFs are probably heated locally and can be produced by magnetic reconnection in the lower atmosphere \citep{machado1989radiative,li1997magnetic}. Apart from local heating, the possible heating mechanisms for type II WLFs also include the bombardment of energetic electron beams with quite a large spectral index and energy transport through Alfv\'{e}n waves. However, the problems such as the energy deposition process in WLFs are not fully solved yet. Until now, many heating mechanisms have been proposed to describe the energy deposition in the temperature minimum region (TMR) and the upper photosphere, which include heating by a non-thermal electron beam (\citealt{hudson1972thick}; \citealt{aboudarham1986energy}), a proton beam \citep{machado1978}, dissipation of Alfv\'{e}n waves (\citealt{emslie1982temperature}; \citealt{fletcher2008impulsive}) and chromospheric condensation \citep{machado1989radiative,gan1994atmospheric}. In this paper, we focus on the non-thermal electron beam heating mechanism and use this model to simulate the WLFs. The rationality for this lies in the well spatially and temporally correlated continuum enhancement and HXR emission reported in previous observations of WLFs \citep{watanabe2010g,krucker2015co,kuhar2016correlation}. In fact, the quantitative relationship between the non-thermal electrons and white-light emission has been investigated in a number of studies from the observational aspect \citep{chen2005relationship,fletcher2007trace,kuhar2016correlation,watanabe2017characteristics}. Recently, \citet{lee2017irisa} studied an X1.6 flare in AR 12192 on 2014 October 22. They compared the energy flux from non-thermal electrons with the dissipated energy estimated from the Mg {\footnotesize{II}} triplet. Their result indicate that the non-thermal electrons could produce the continuum enhancement directly. By studying an X1.8-class ﬂare with a strong white-light emission on 2012 October 23, \citet{watanabe2020white} believed that the accelerated electrons could be the source of the white-light emission. These works increase the rationality of our simulations.

Since WLFs are believed to originate from a deep layer in the solar atmosphere (e.g., \citealt{machado1974analysis,rust1975analysis}), which is characterized by low temperature, high density and partial ionization, obtaining the physical parameters of such a flare atmosphere from the spectral observations is a very complex problem. To uncover the physical mechanisms under WLFs, we need to resort to sophisticated simulations of the flare atmosphere by including as many as possible the key physical processes like gas dynamics and radiative transfer. With such simulations, we can learn how the continuum emission, as well as the spectral lines, respond to atmospheric heating especially by an electron beam, and what spectral characteristics can be used in order to diagnose the parameters of the electron beam.

In recent years, radiative hydrodynamic simulations have been an important tool to study the solar and stellar WLFs. Via RADYN simulations, \citet{allred2005radiative} reported that the line and continuum emissions can show remarkable increases in the optical and UV bands during both moderate and strong flares. \citet{cheng2010radiative} used RADYN to successfully reproduce the observational white-light enhancements except for the most energetic WLFs. An M-dwarf WLF was modeled with RADYN by \citet{kowalski2015new}, in which excessive optical and near UV continuum emission was produced from a hot and dense chromospheric condensation. They also reproduced self-consistently the observed Balmer jump ratio and color temperature for the first time. However, these results came from electron beams with a very high flux of $10^{13}\ \rm{erg}\,\rm{cm}^{-2}\,\rm{s}^{-1}$ for dMe flares,  which is very different from the typical value of $10^{11}\ \rm{erg}\,\rm{cm}^{-2}\,\rm{s}^{-1}$ during solar flares. Quite recently, \citet{prochazka2018reproducing, prochazka2019hydrogen} did some simulations using the RADYN code with the concentration on type II solar WLFs, to reproduce the observed X1.0 solar WLF on 2014 June 11 and study the relationship between hydrogen radiative losses and the low-energy cutoff of the electron beams. The authors pointed out that the absence of significant Lyman excess emission with high low-energy cutoff and low fluxes was a prominent spectral feature for type II WLFs. As for the hydrogen lines, Ly$\alpha$ or H$\alpha$, a strong relationship between their emission increase and non-thermal electron beams has been reported by \citet{allred2005radiative} and \citet{prochazka2019hydrogen}. In addition, \citet{kowalski2017hydrogen} has noted the importance of high fluxes of non-thermal electron beam in the broadening of the Balmer lines.

In this paper, we carry out RADYN simulations to study the response of the flaring atmosphere to non-thermal electron beams which has been proved to play an important role in increasing the continuum emission \citep{allred2005radiative, hudson2006white, jess2008do, cheng2010radiative, prochazka2018reproducing}. The continua at 3600\angstrom\ and 4250\angstrom, i.e., below and above the Balmer jump at 3646\angstrom\  respectively, are calculated here. We also include the H$\alpha$ and Ly$\alpha$ lines that are formed ranging from the photosphere to chromosphere and even the transition region \citep{vernazza1981structure} to gain a more comprehensive understanding of the features of the flare atmosphere. Note that the continua at 3600\angstrom\ and 4250\angstrom\ are two wavelengths that are used in the Optical and Near-infrared Solar Eruption Tracer (ONSET; \citealt{fang2013new}) observations. Moreover, the continuum at 3600\angstrom\ and the Ly$\alpha$ line are two available observation wavebands for the future LST/ASO-S telescope (Lyman-alpha Solar Telescope/Advanced Space-based Solar Observatory, \citealt{li2019lyman}). Therefore, the theoretical calculation can be used as a reference for analyzing future WLF observations. 

Here we briefly introduce the numerical method used in our study in Section \ref{sect:method}. In Section \ref{sect:result}, we present the simulation results of the continua at 3600\angstrom\ and 4250\angstrom\ together with the H$\alpha$ and Ly$\alpha$ lines in response to varying the electron beams. Finally, the discussions and conclusions are given in Section \ref{sect:discussion}.

\section{Numerical Method}
\label{sect:method}

\subsection{Flare Simulation}

The one-dimensional (1D) non-local thermodynamic equilibrium (non-LTE) radiative hydrodynamics code, RADYN, uses a self-adaptive grid \citep{dorfi1987simple} to implicitly solve the hydrodynamic and radiative transfer equations \citep{carlsson1992non,carlsson1995does,carlsson1997formation,carlsson2002dynamic}. More details about the recent implements can be found in \citet{allred2015unified}. In this work, we employ this code to calculate the atmospheric response to non-thermal electron beams with different parameters during WLFs, which has been proven practicable and reasonable in the studies mentioned above. 

We assume that the energy is injected at the top of the flare loop in the form of a non-thermal electron beam. And the loop extends for a length of 10\,Mm from photosphere to corona with a quarter-circular structure. Then the electrons propagate downward and deposit their energy and momentum in the lower atmosphere to heat and ionize the ambient plasma. The heating rate of the electron beam is quantified by solving the Fokker-Planck equation \citep{mctiernan1990behavior, allred2015unified}. Atoms which show great significance to the chromospheric energy balance are treated in non-LTE. More specifically, hydrogen and Ca {\footnotesize{II}} atoms are modeled with six levels including a continuum level, on the other hand, helium atoms are modeled with nine levels including continuum. All radiative transitions between these levels are calculated in detail and assumed to be under complete frequency redistribution (CRD) to simplify the calculation. Other atoms are treated in LTE and included in the calculation by means of the Uppsala opacity package \citep{gustafsson1973}. The return current is not included in the simulations for the following two reasons. (1) The return current only shows a great impact on coronal heating \citep{allred2015unified} and does not significantly affect the response of continua as well as spectral lines that are formed in the chromosphere or photosphere. (2) The non-thermal electron beam heating may overweight the return current heating considerably in the chromosphere. 
An extra cooling term due to thermal bremsstrahlung and metal collisionally-excited transitions is added into the internal energy conservation equation to include the optically thin radiative cooling. This cooling term is obtained from CHIANTI atomic database \citep{dere1997chianti,dere2019chianti}.

\begin{figure}[h]
   \centering
   \includegraphics[width=\textwidth, angle=0]{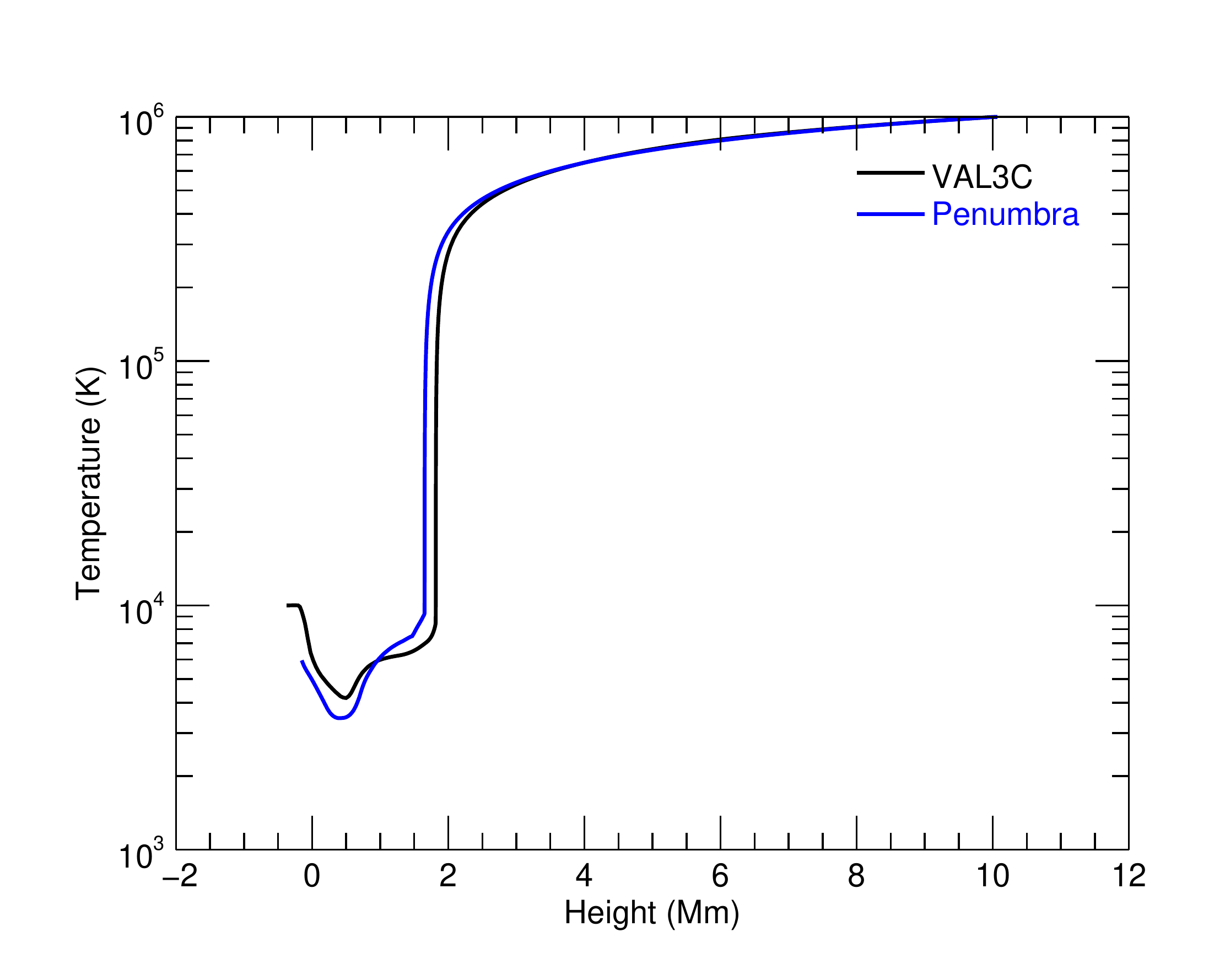}
   \caption{Temperature distribution of the initial atmosphere after relaxation. The black curve is for the quiet-Sun (VAL3C) and the blue curve is for the penumbra. }
   \label{Fig1}
   \end{figure}

Taking into account the fact that the solar WLFs often arise in or near the sunspots, we adopt the penumbral atmosphere model as the initial atmosphere \citep{hong2018non}, which is constructed based on the semi-empirical model of \citet{ding1989semi-empirical}. For comparison, we also employ a quiet-Sun model based on the VAL3C atmosphere \citep{vernazza1981structure} which has been used in some previous works \citep{hong2017radyn,hong2018non,hong2019response}. 
The main difference between these two model atmospheres is that the penumbral model features a lower temperature in the lower atmosphere, whereas it features a higher temperature in the higher atmosphere, which can be seen in Figure~\ref{Fig1}.

In our simulations, we assume that the initial atmospheres are heated by electron beams during a flare. The electron beams are assumed to have a power-law distribution with two different spectral indices of $\rm \delta = 3$ and 5 and a low-energy cutoff of $E_c=25$\,keV. These parameters are selected based on the WLF studies by \citet{kuhar2016correlation}, \citet{fletcher2007trace}, and \citet{watanabe2010g}. The former one found that the observed spectral indices in 43 WLFs are mainly between 2.5 and 5.5, and the latter two demonstrate that an electron beam with a low-energy cutoff around 20--30\,keV is sufficient to power the flare WL enhancement. The energy flux of the beam follows a triangular function over time, with a total duration of 20\,s. We consider five different average energy fluxes (half of the maxima), $F\;$=$\;10^{9}$, $\;3\times10^9, \;10^{10}, \;3\times10^{10}$, $\;5\times10^{10}$~erg\,cm$^{-2}$\,s$^{-1}$, which are denoted as F9, 3F9, F10, 3F10 and 5F10, respectively. Such a range of energy fluxes represents variation from a relatively weak flare to a powerful one\citep{abbett1999dynamic}. 
We run all the cases for two different spectral indices, five  different energy fluxes, as well as two different initial atmospheres. This means that we have 20 different sets of parameter combinations. Note that here we only consider a disk position of $\mu=0.953$ (around the disk center). The model parameters of all 20 cases are listed in Table~\ref{Tab:para}. Each case is run for 20\,s and the simulation snapshots are saved every 0.1\,s. It should be mentioned that, performing radiative hydrodynamic simulations of flares even for 1D model is computationally challenging and time-consuming, in particular with a higher heating rate and a larger spectral index. Therefore, in our work, we did not finish the computation in some cases, i.e., 3F10d5Q, 5F10d5Q, 3F10d5P, and 5F10d5P, which can be seen in all of the result figures.

\begin{table}
\begin{center}
\caption[]{ List of Parameters of Our Simulations }\label{Tab:para}
 \begin{tabular}{cccc}
  \hline\noalign{\smallskip}
Label & $F$ $\rm
(erg~s^{-1}~cm^{-2})$ & $\delta$ & Initial Atmosphere    \\
  \hline\noalign{\smallskip}
F9d3Q & $10^{9}$ & 3 & Quiet Sun \\
3F9d3Q & $3\times10^{9}$ & 3 & Quiet Sun \\
F10d3Q & $10^{10}$  & 3 & Quiet Sun \\
3F10d3Q & $3\times10^{10}$  & 3 & Quiet Sun \\
5F10d3Q & $5\times10^{10}$  & 3 & Quiet Sun \\
F9d5Q & $10^{9}$  & 5 & Quiet Sun \\
3F9d5Q & $3\times10^{9}$  & 5 & Quiet Sun \\
F10d5Q & $10^{10}$  & 5 & Quiet Sun \\
3F10d5Q & $3\times10^{10}$  & 5 & Quiet Sun \\
5F10d5Q & $5\times10^{10}$  & 5 & Quiet Sun \\
F9d3P & $10^{9}$  & 3 & Penumbra \\
3F9d3P & $3\times10^{9}$  & 3 & Penumbra \\
F10d3P & $10^{10}$  & 3 & Penumbra \\
3F10d3P & $3\times10^{10}$  & 3 & Penumbra \\
5F10d3P & $5\times10^{10}$  & 3 & Penumbra \\
F9d5P & $10^{9}$  & 5 & Penumbra \\
3F9d5P & $3\times10^{9}$  & 5 & Penumbra \\
F10d5P & $10^{10}$  & 5 & Penumbra \\
3F10d5P & $3\times10^{10}$  & 5 & Penumbra \\
5F10d5P & $5\times10^{10}$  & 5 & Penumbra \\
  \noalign{\smallskip}\hline
\end{tabular}
\end{center}
\end{table}

\subsection{Calculation of the Ly{$\alpha$} Line}

RADYN assumes all transitions to be treated in CRD. However, for Ly$\alpha$ line, the partial frequency redistribution (PRD) effects are crucial since the simplified CRD calculation can notably overestimate the radiative losses in strong resonance lines, which causes a false atmospheric temperature structure. Examples can be found in \citet{ulmschneider1987acoustic} and \citet{uitenbroek2002effect} for Mg {\footnotesize{II}} k or Ca {\footnotesize{II}} H \& K cases. The effects of PRD on the Ly$\alpha$ line have been shown in \citet{hong2019response} and \citet{kerr2019a}. Motivated by these two studies, we employ the radiative transfer code RH \citep{uitenbroek2001multilevel, pereira2015rh} to calculate the Ly$\alpha$ line profiles. Note that RH assumes the atmosphere to be in statistical equilibrium, which means it does not satisfy the condition of the highly dynamic impulsive phase in WLFs perfectly \citep{carlsson2002dynamic}. However, the effect of this assumption is not as efficacious as PRD, as discussed by \citet{kerr2019a,kerr2019b}. Here, for each case, we take snapshots every 0.1\,s from RADYN outputs which contain hydrogen level populations and electron density as inputs to the RH code to calculate the Ly$\alpha$ line profiles.

\section{Simulation results}
\label{sect:result}

\subsection{Continuum Contrast}
\label{sect:continuum}

\begin{figure}[h]
   \centering
   \includegraphics[width=\textwidth, angle=0]{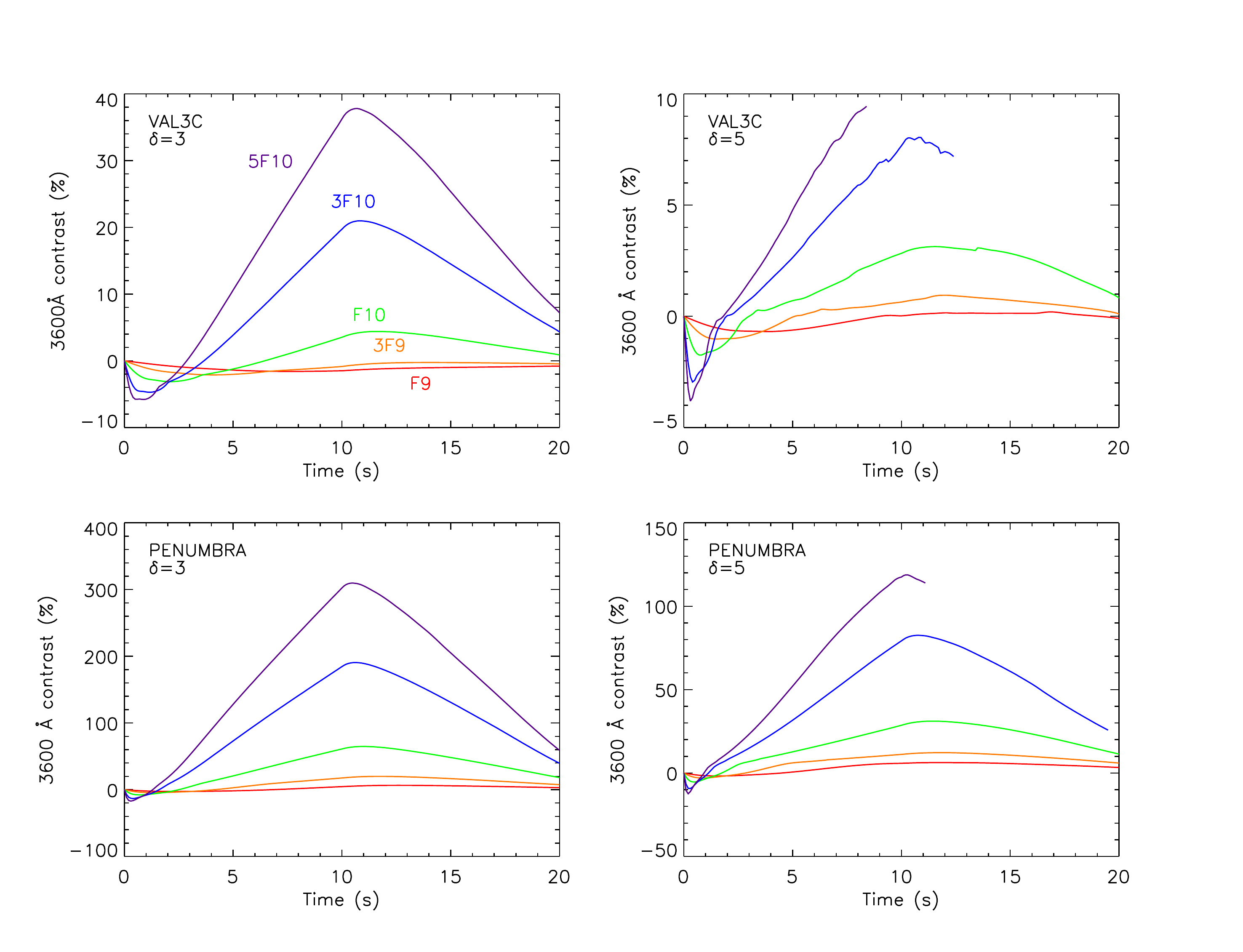}
   \caption{Time evolution of the continuum contrast at 3600{\,\AA} with various combinations of initial atmosphere, average energy flux $F$ and spectral index $\delta$. }
   \label{Fig2}
   \end{figure}

\begin{figure}[h]
   \centering
   \includegraphics[width=\textwidth, angle=0]{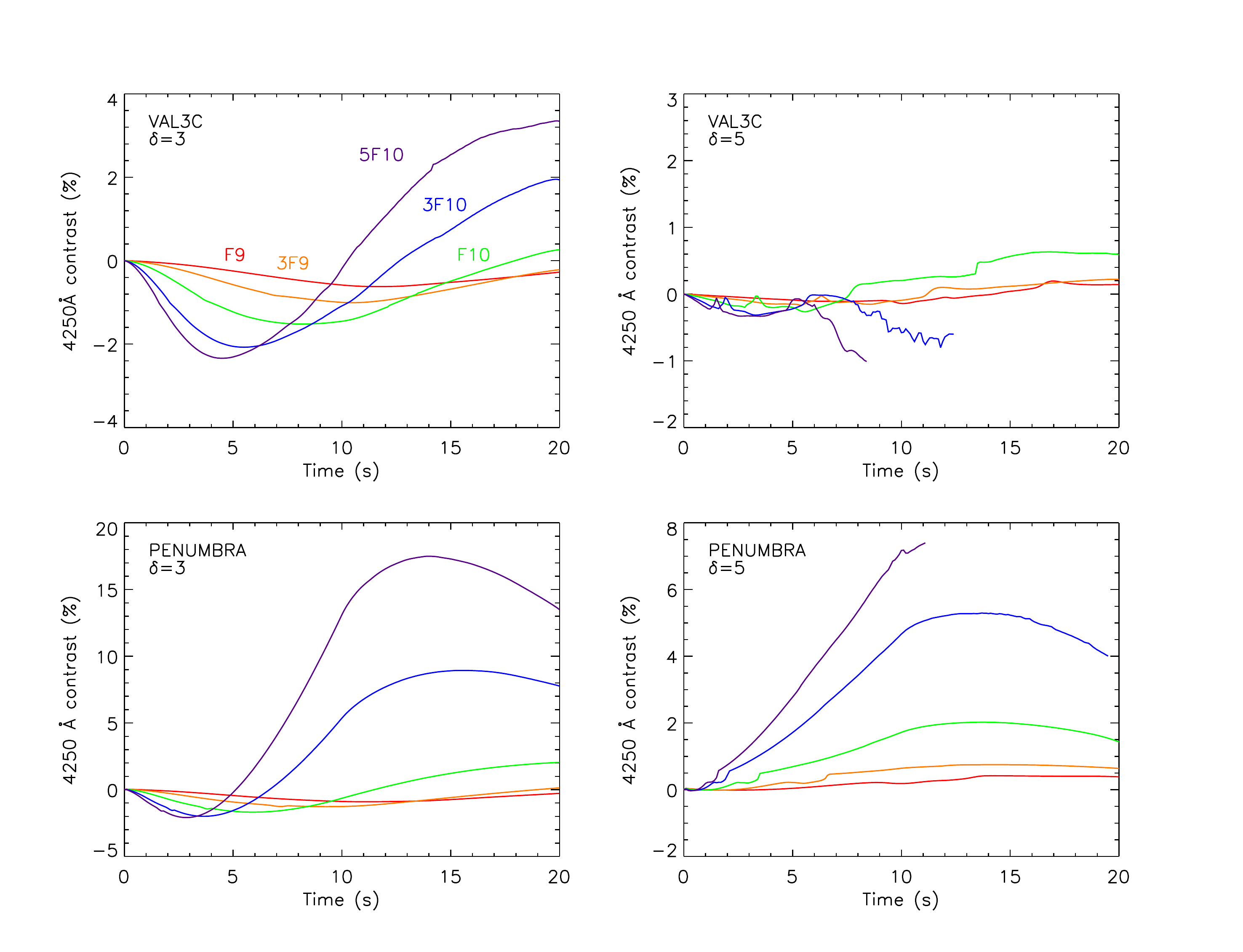}
   \caption{Same as Figure~\ref{Fig2} but for the continuum contrast at 4250{\,\AA}. }
   \label{Fig3}
   \end{figure}

We calculate the continuum contrast for different models, which is defined as $\Delta I/I_0 = (I - I_0)/I_0$ where $I$ and $I_0$ are the continuum intensities for the flaring and pre-flare atmospheres, respectively. Figures~\ref{Fig2} and \ref{Fig3} present time evolutions of the continuum contrasts at 3600\angstrom\ and 4250\angstrom\ with various combinations of initial atmosphere, average energy flux $F$ and spectral index $\delta$. We can see that both of the continuum emissions are enhanced as the heating sets in. The enhancement at both wavelengths is greater with a larger flux of the injected electron beam, a smaller spectral index and a penumbral atmosphere. The difference in the enhancement is particularly obvious when the initial atmosphere and electron flux changes. For instance, for cases with $\delta = 3$ and quiet-Sun atmosphere, the highest two fluxes result in the continuum contrast at 3600\angstrom\ larger than 20\%. However, the lowest two fluxes hardly show any enhancement during the heating. The effect of initial atmosphere can be seen when cases 3F10d3Q and 3F10d3P are compared. The former exhibits a 3600\angstrom\ contrast a bit larger than 20\%, while the latter shows a contrast near 200\%. We also find that the enhancement of continuum at 4250\angstrom\ is less than that of continuum at 3600\angstrom. In terms of the percentage of enhancement, the former is one order of magnitude smaller than the latter. For example, for the case 5F10d3P, the continuum enhancement at 4250\angstrom\ is about 17\%, while for 3600\angstrom\ this value is around 300\%.

It is also seen that at the beginning of the heating, the continua at both 3600{\,\AA} and 4250{\,\AA} have a dimming with durations and strengths varying with model parameters. The dimming timescale of 4250{\,\AA} is longer than that of 3600{\,\AA}. This is obviously presented in the cases with a spectral index of 3 (Figures~\ref{Fig2} and \ref{Fig3}). For these cases, the typical duration of dimming at 3600{\,\AA} is less than 3\,s, while for 4250{\,\AA}, almost all the dimming last for at least 5\,s. 
Cases with higher electron beam fluxes display a shorter dimming, and the strength of the dimming shows a negative correlation with its duration. Cases with a quiet-Sun initial atmosphere exhibit a longer duration than those with a penumbral atmosphere. Since the dimming time can be very long in some cases with low energy fluxes (sometimes even more than 10\,s, for example, the case F10d3Q), the light curve is still increasing at the end of heating.

\begin{figure}[h]
   \centering
   \includegraphics[width=\textwidth, angle=0]{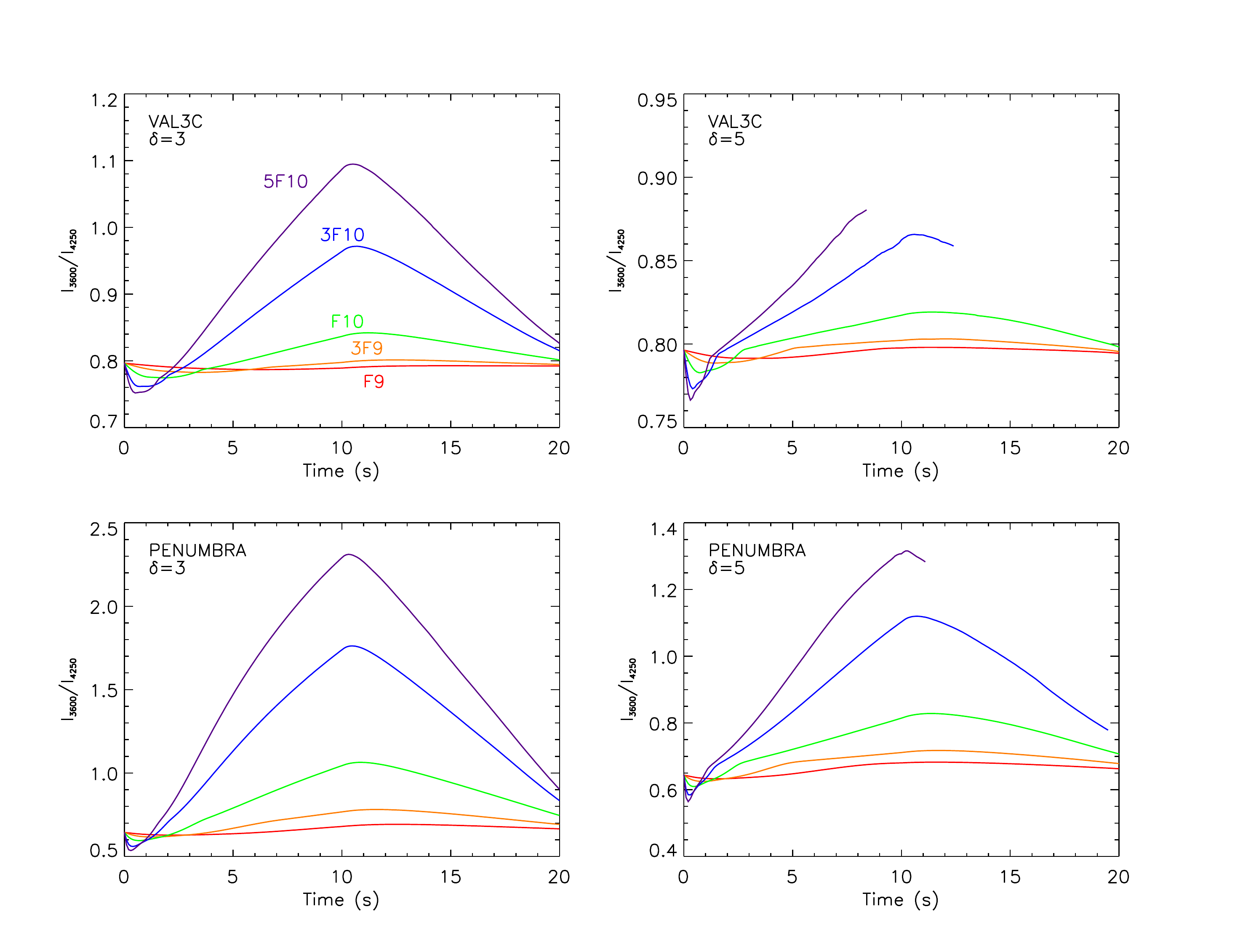}
   \caption{Time evolution of the ratio of I$_{3600}$/I$_{4250}$ with various combinations of initial atmosphere, average energy flux $F$ and spectral index $\delta$. }
   \label{Fig4}
   \end{figure}

 In addition, we calculate the ratio of the intensity at 3600{\,\AA} to that at 4250{\,\AA} (i.e., $I_{3600}/I_{4250}$, as shown in Figure~\ref{Fig4}), which could reflect the intensity variation of below and above the Balmer jump to some extent.  
It is seen that this ratio is less than unity before heating. With the heating going on, the percentage of enhancement at 3600{\,\AA} is one order of magnitude higher than that of 4250{\,\AA}, so the curve for this ratio has a shape similar to that for the contrast at 3600{\,\AA}, with a decrease at first and then an increase to the peak value at over 10\,s after heating. However, if using a penumbral initial atmosphere, a higher energy flux and a smaller spectral index, we can get a larger ratio of $I_{3600}/I_{4250}$. The maximum ratio in all cases is about 2.3 for the 5F10d3P run.  It should be mentioned that the ratio can be more easily larger than unity when the penumbral initial atmosphere is applied.

\subsection{The H{$\alpha$} and Ly{$\alpha$} Integrated Intensity} 
\label{sect:halpha intensity}

\begin{figure}[h]
   \centering
   \includegraphics[width=\textwidth, angle=0]{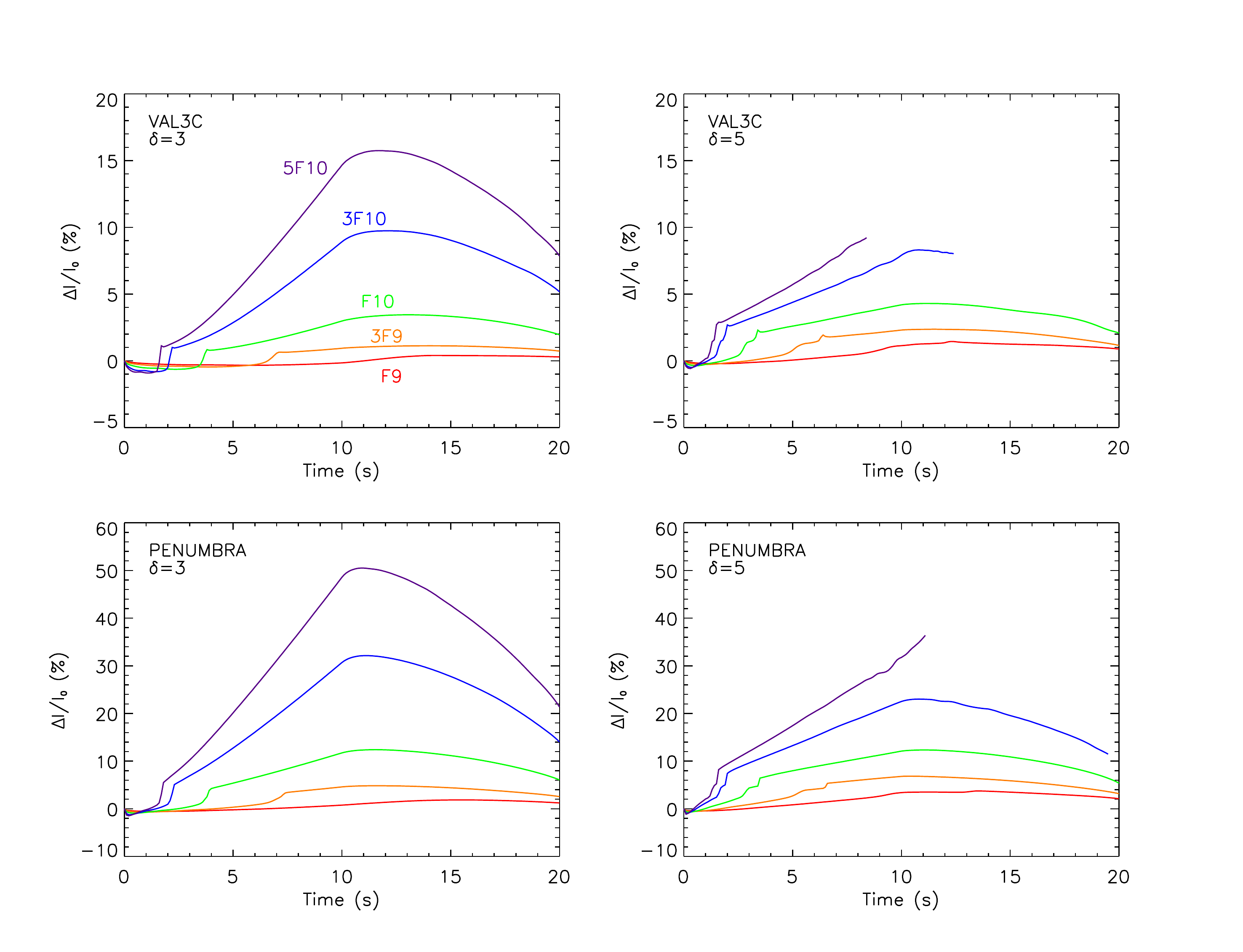}
   \caption{Time evolution of the integrated intensity contrast of the H$\alpha$ line with various combinations of initial atmosphere, average energy flux $F$ and spectral index $\delta$. The integration range is from 6520.5{\,\AA} to 6609.1{\,\AA}. }
   \label{Fig5}
   \end{figure}

\begin{figure}[h]
   \centering
   \includegraphics[width=\textwidth, angle=0]{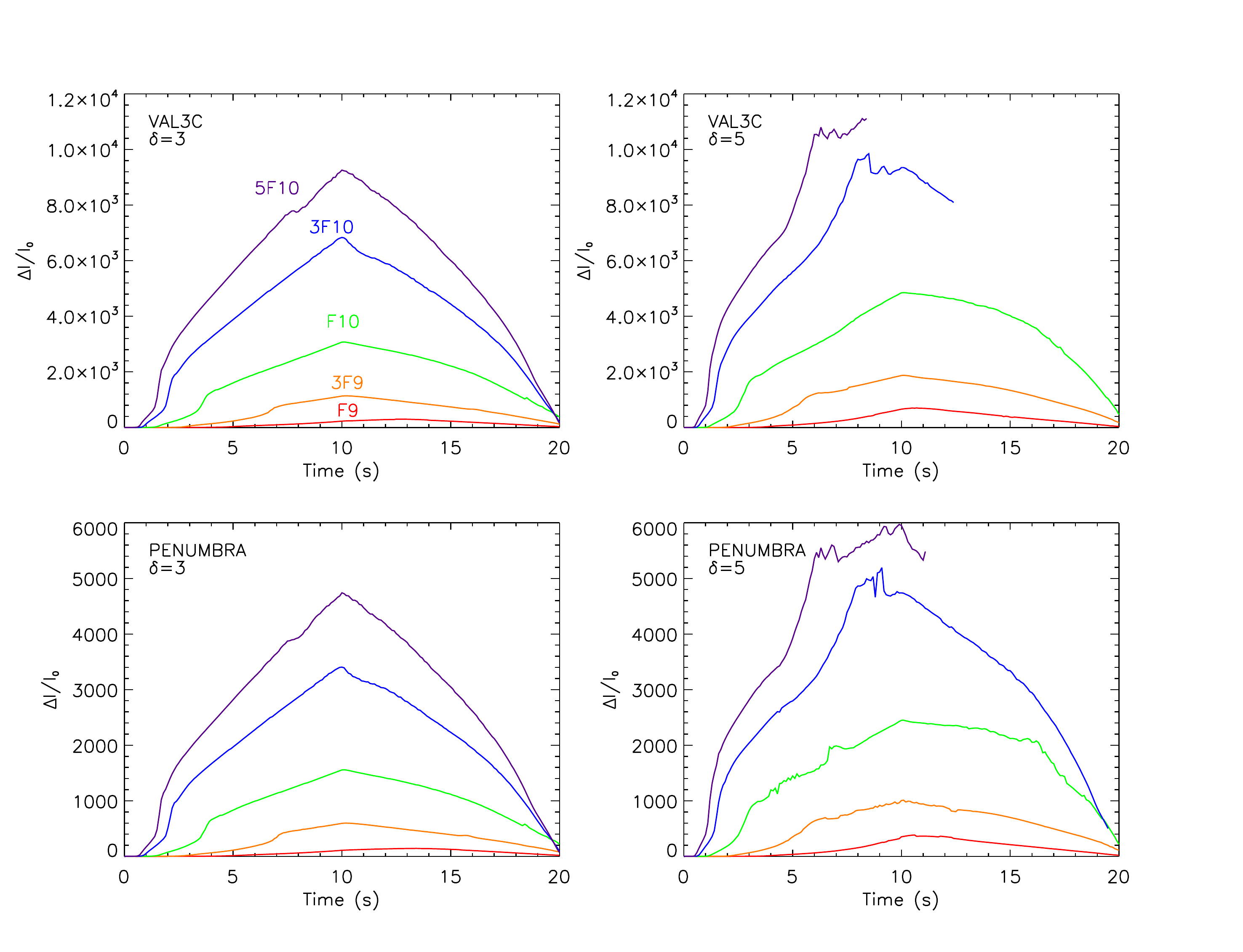}
   \caption{Same as Figure~\ref{Fig5} but for the Ly$\alpha$ line with an integration range from 1215.0{\,\AA} to 1216.5{\,\AA}. }
   \label{Fig6}
   \end{figure}

Figures~\ref{Fig5} and \ref{Fig6} show the time evolution of the integrated intensity contrasts of the H$\alpha$ and Ly$\alpha$ lines for different models, which have a similar definition with the continuum contrast. From these two figures, we can find that for quiet-Sun and penumbral initial atmospheres, the integrated intensity contrasts for both H$\alpha$ and Ly$\alpha$ lines can reflect the effect of electron beam bombardment, since all the curves exhibit an increase in the first 10\,s. As the electron beam flux increases, the emission of the lines also increases. However, the H$\alpha$ and Ly$\alpha$ lines do not peak synchronously. For the H$\alpha$ line, the peak time in all cases is apparently later than 10\,s, which is the peak time of the electron energy flux. While for the Ly$\alpha$ line, the intensity peaks at the peak time of the electron energy flux.

The initial atmosphere has also a significant influence on the results. We can notice that the enhancement in the Ly$\alpha$ intensity is larger when the quiet-Sun model is used (Figure~\ref{Fig6}), whereas for the H$\alpha$ line the penumbral model displays a stronger enhancement of the line intensity (Figure~\ref{Fig5}). The peak values of the contrast for the H$\alpha$ line in the penumbral atmosphere are basically three times larger than the peak values in the quiet-Sun atmosphere. And the peak values of the Ly$\alpha$ line in the quiet-Sun atmosphere are roughly two times greater than the peak values in the penumbral atmosphere. 

We can also check the effects of spectral index. By comparing the left and right panels of Figure~\ref{Fig6}, we can find that the Ly$\alpha$ line shows a greater enhancement when the spectral index is larger, which corresponds to a softer energy spectrum of the non-thermal electrons. However, as shown in Figure~\ref{Fig5}, the H$\alpha$ line displays a contrary mode: a larger enhancement can be achieved when a harder electron spectrum is applied.

From Figure~\ref{Fig5}, we can also see that the H$\alpha$ line exhibits a dimming when the injection of non-thermal electrons begins, which is consistent with the pattern shown in continuum emission. Its dimming timescale is basically the same as the one for the continuum at 3600{\,\AA}. When the spectral index gets smaller, the dimming is more significant and lasts for a longer time. A larger electron beam flux is likely to cause a more evident dimming with a shorter duration time. But from Figure~\ref{Fig6}, it is seen that no dimming can be found in the Ly$\alpha$ line. This difference is explained in Section~\ref{sect:discussion}.

The electron beam with a larger flux would cause an earlier enhancement and a larger peak value. Compared with the H$\alpha$ line, the enhancement of the Ly$\alpha$ line is more significant. Taking the two cases 5F10d3Q and 5F10d3P as examples, the contrast can reach 4700 times in the penumbral atmosphere and 9000 times in the quiet-Sun atmosphere for the Ly$\alpha$ line; while for the H$\alpha$ line, the largest contrast would be about 50\%. It is also noteworthy that, when the electron flux is 5F10, there is a small ``platform" whose slope becomes smaller or even negative before reaching the peak value (see Figure~\ref{Fig6}). This ``platform" appears at about 8\,s and lasts for less than 1\,s when the spectral index is 3. When the spectral index is 5, it appears at about 6\,s and lasts for about 2\,s. This result is consistent with the intensity dip produced by \cite{hong2019response}, in which the occurrence of the ``platform" is supposed to be related with the line asymmetry change.

\subsection{The H{$\alpha$} and Ly{$\alpha$} Line Profiles}

\begin{figure}[h]
   \centering
   \includegraphics[width=\textwidth, angle=0]{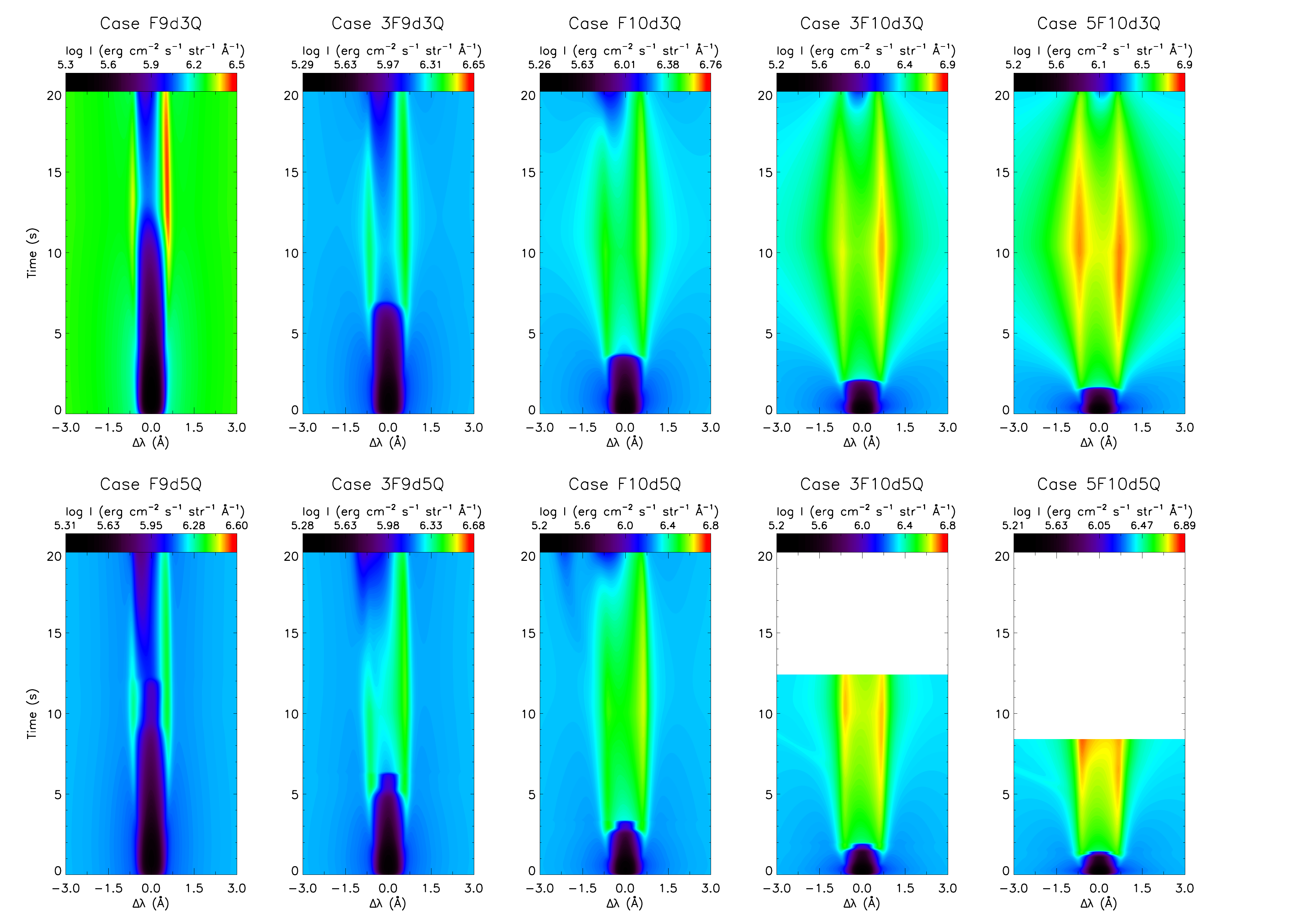}
   \caption{Time evolution of the H$\alpha$ line profiles with the quiet-Sun initial atmosphere and spectral index $\delta$ $\;$=$\;$3 (top panels) and 5 (bottom panels).}
   \label{Fig7}
   \end{figure}

\begin{figure}[h]
   \centering
   \includegraphics[width=\textwidth, angle=0]{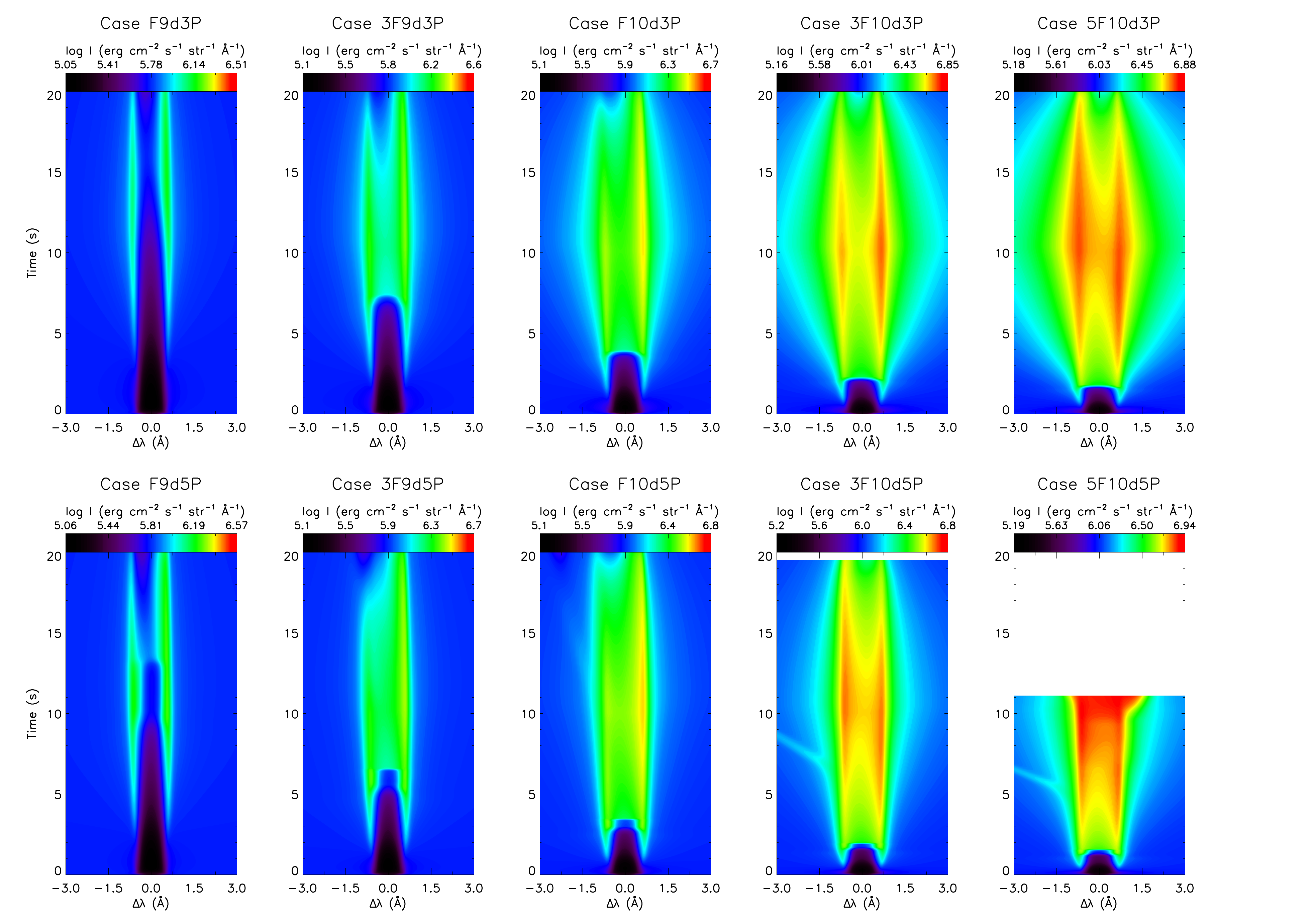}
   \caption{Same as Figure~\ref{Fig7} but with the penumbral initial atmosphere.}
   \label{Fig8}
   
   \end{figure}

\begin{figure}[h]
   \centering
   \includegraphics[width=\textwidth, angle=0]{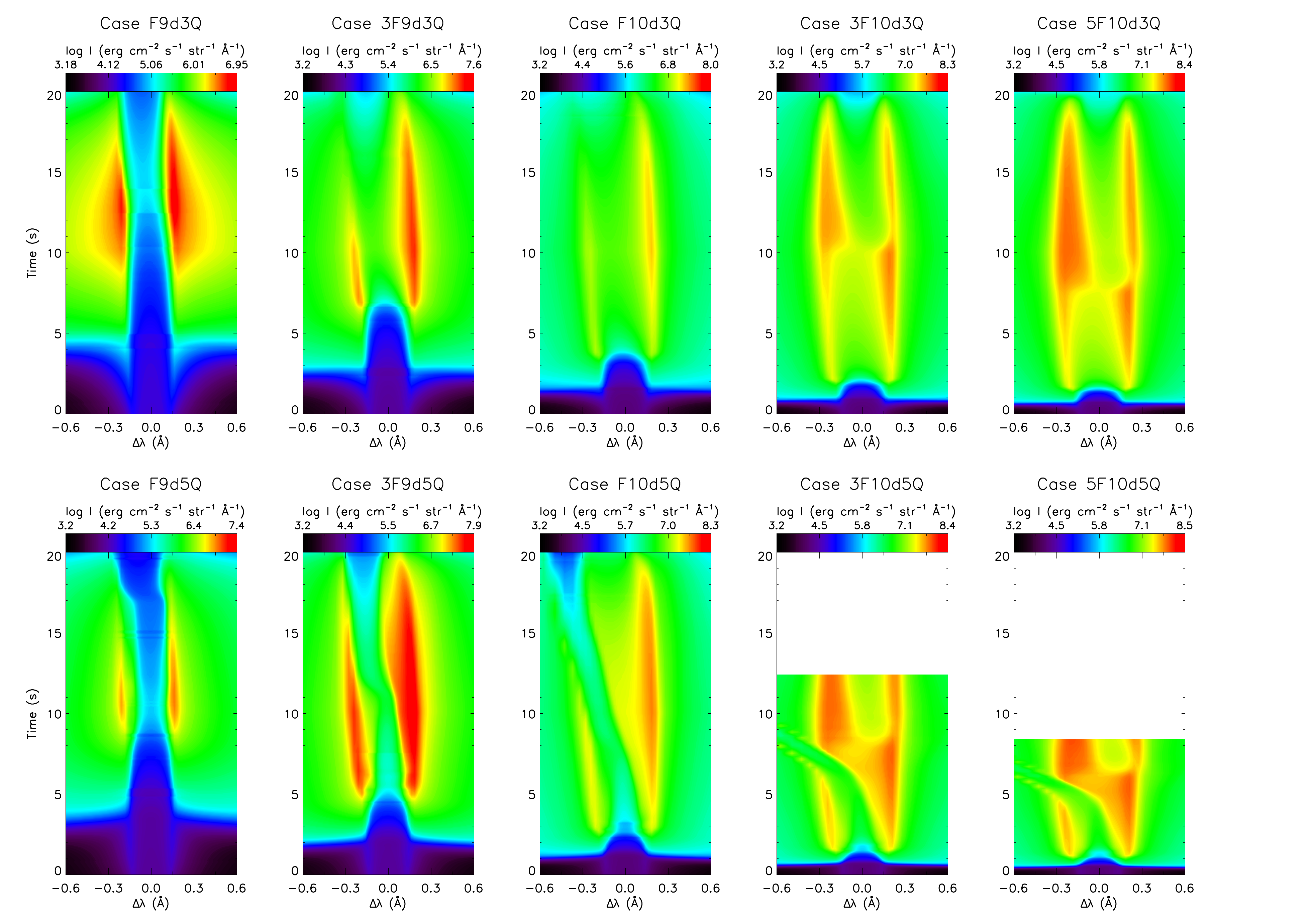}
   \caption{Same as Figure~\ref{Fig7} but for the Ly$\alpha$ line. }
   \label{Fig9}
   \end{figure}

\begin{figure}[h]
   \centering
   \includegraphics[width=\textwidth, angle=0]{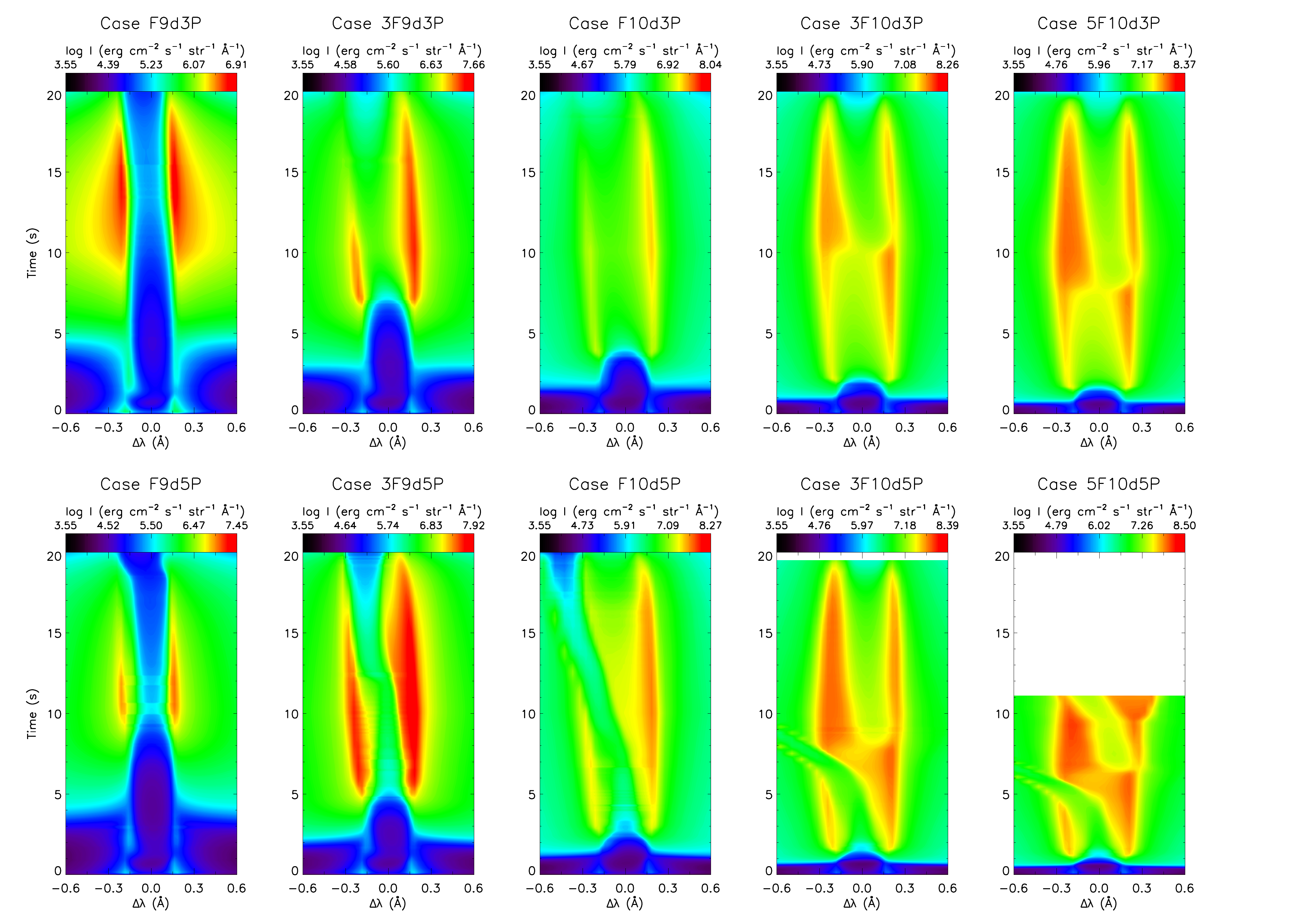}
   \caption{Same as Figure~\ref{Fig8} but for the Ly$\alpha$ line.}
   \label{Fig10}
   \end{figure}

We plot the time evolution of the H$\alpha$ and Ly$\alpha$ line profiles for different models in Figures~\ref{Fig7}--\ref{Fig8} and Figures \ref{Fig9}--\ref{Fig10}, respectively. By comparing the profiles of the H$\alpha$ and Ly$\alpha$ lines in different models, it can be found that the electron beam with different parameters (i.e., the average energy flux and spectral index) plays a more effective role in the enhancement of the lines than the initial atmosphere, which is a little different from the enhancement of the continuum emission as described in Section \ref{sect:continuum}.

      
In the whole heating process of the flare atmosphere, there is always a central reversal in the Ly$\alpha$ line profile. For the case with an average energy flux F10 or below, as the heating goes on, the Ly$\alpha$ line center gradually shifts towards the blue wing, and therefore the spectral line profile presents  a red asymmetry. This indicates that the Ly$\alpha$ line center is formed at the height where an upward velocity exists (i.e., a chromospheric evaporation layer; see more details in \citealt{hong2019response}). For the cases with average energy fluxes 3F10 and 5F10, the Ly$\alpha$ line center displays a blueshift in the first few seconds of heating, resulting in a red asymmetry.
However, when the heating continues to a certain extent, at about 10\,s and 8\,s for 3F10 and 5F10 cases, respectively, the line center changes from blueshift to redshift. At the same time, the line profile changes from a red asymmetry to a blue asymmetry. 
This implies that at this moment, the Ly$\alpha$ line center is formed in a layer where a downflow exists, i.e., a chromospheric condensation layer. To be more specific, the formation height of the Ly$\alpha$ line center decreases to a region where a condensation appears, at least in these two cases \citep{hong2019response}. After the asymmetry change, the Ly$\alpha$ line center gradually moves back to the stationary position.

For the H$\alpha$ line, before the heating, the profile exhibits a central absorption. With the heating going on, an emission appears in the H$\alpha$ line wing. For the case of an intense heating, i.e., average energy flux of 5F10 and spectral index of 3, the H$\alpha$ line profile also shows a change from red asymmetry to blue asymmetry, but such a transition occurs later than Ly$\alpha$ under the same flux of 5F10. The former occurs at about 10\,s after the beginning of heating, in comparison with the latter at about 8\,s.

\section{Discussions and conclusions}
\label{sect:discussion}

In this paper, we carry out radiative hydrodynamic simulations by using RADYN to investigate the response of the continua at 3600{\,\AA} and 4250{\,\AA} together with the H$\alpha$ and Ly$\alpha$ lines by varying the initial atmospheres and the parameters of the electron beam as the main heating source. We find that the electron beam flux is a very important factor to influence the emission excess of continua and hydrogen lines, especially when it is larger than F10. 
An important finding is that the electron beam flux is found to influence the enhancement of H$\alpha$, Ly$\alpha$, and continuum in a similar way, whereas the other two factors, the initial atmosphere and the spectral index, show different impacts on the enhancement. For the H$\alpha$ line as well as the continua at 3600{\,\AA} and 4250{\,\AA}, a penumbral initial atmosphere and a smaller spectral index can cause a greater emission enhancement, whereas the Ly$\alpha$ line is more favourably enhanced in a quiet-Sun initial atmosphere and a larger spectral index. 
In addition, it is found that the Ly$\alpha$ line intensity peaks nearly at the same time as the heating function, while the H$\alpha$ line and the continua at 3600{\,\AA} and 4250{\,\AA} reach their peak emissions later than that. 

The different responses of the Ly$\alpha$ line with the H$\alpha$ line as well as the continua at 3600{\,\AA} and 4250{\,\AA} could be explained by their different formation heights and also the initial atmosphere structure. It is known that the Ly$\alpha$ line is formed in the mid-chromosphere to the low transition region \citep{vernazza1981structure} where the quiet-Sun initial atmosphere has a lower temperature, while the H$\alpha$ line and the continua at 3600{\,\AA} and 4250{\,\AA} are formed in the chromosphere to the photosphere \citep{vernazza1981structure} where the penumbral initial atmosphere features a lower temperature in principle. A lower temperature structure should yield a greater emission enhancement. When we consider the effects of spectral index of the electron beam, this is mainly due to the different depths that electron beams with different spectral indices can penetrate. As reported by \cite{allred2015unified}, a harder electron beam can deposit its energy into a deeper layer which preferably enhances spectral lines forming in the lower atmosphere like H$\alpha$, while a softer electron beam deposits its energy into a relatively higher layer that is more favourable for the Ly$\alpha$ line emission. 

To the best of our knowledge, this is the first time to calculate the response of the continuum at 3600{\,\AA} which has already been and will also be used in the WLF observations. In particular, we calculate the response of the atmosphere through the photosphere to the low transition region in a same electron beam heating regime. 
The results presented in this work can improve our understanding on the heating mechanisms and energy propagation in WLFs. It is found that the Ly$\alpha$ intensity peaks nearly at the same time as the electron beam heating function in our 20 models. This may suggest that the Ly$\alpha$ emission is directly related to the non-thermal electron beams. In other words, the electron beams deposit their energy mostly in the chromosphere and heat the local plasma, then we see the evident Ly$\alpha$ emission. In the following, the continuum at 3600{\,\AA} as well as the H$\alpha$ line show an emission peak, and finally the continuum at 4250{\,\AA} reaches its peak emission. This time sequence may indicate that the deposited energy is transported to the lower atmosphere and heats the local plasma successively. Therefore, we can track the energy propagation from the observed light curves at these wavelengths. It should be noted that the cases with a very soft beam or a thermal heating could be different in the processes of energy deposition and propagation (based on a work in preparation).

Furthermore, a penumbral initial atmosphere is employed for the first time to study the response of the continuum emission in the present work, which is proved to be more favourable for the WL emission enhancement when compared with the quiet-Sun initial atmosphere. We notice that \citet{cheng2010radiative} had employed a sunspot initial atmosphere in the WLF simulations and the simulations can explain most of the observed WL emissions. As for the semi-empirical penumbral atmosphere from \citet{ding1989semi-empirical}, it was used to generate the initial atmosphere for the calculations of the Fe {\footnotesize{I}} 6173 \angstrom\ line in a flaring atmosphere by \citet{hong2018non}, but has not been employed for the WLF simulations using RADYN code. From our results, it can be found that the initial atmospheres based on different models do cause different simulation outcomes. We find that for both 3600{\,\AA} and 4250{\,\AA} the enhancement is stronger when the penumbral atmosphere is used. Therefore, we believe the penumbral atmospheric condition, such as a lower temperature at a lower height, can facilitate the WL emission which originates from the lower atmosphere. In fact, this is consistent with the observational facts that solar WLFs usually occur near sunspots or in sunspots. 

Our results also imply that a continuum enhancement can be formed in small flares, say with an electron beam flux of only strength F9. Therefore, WLFs can be small flares, which is in agreement with the observation results by \citet{hudson2006white}. However, a WL emission cannot be produced if the electron beam heating is not long enough, since there exists a dimming in the continuum at the beginning. 
Note that the continuum dimming caused by electron beams has been exhibited in the simulations of solar and stellar flares \citep{abbett1999dynamic,allred2006radiative} and also in that of Ellerman bombs \citep{hong2017radyn}. Such a dimming is caused by an over-population of the hydrogen excited levels, and thus an enhanced opacity at the corresponding line and continuum, when the atoms are impacted by the non-thermal electrons. In particular, the continua at 3600{\,\AA} and 4250{\,\AA} are mainly contributed by Balmer and Paschen continua, respectively. Therefore, the increases in the populations of the first and second excited states of hydrogen cause the dimming in the continua at 3600{\,\AA} and 4250{\,\AA}, respectively.

The duration of the dimming in the continua depends on the trade-off between recombinations and photoionizations \citep{abbett1999dynamic}. The over-population of the hydrogen excited levels leads to a higher photoionization rate. But as the flare heating goes on, the electron density in the upper chromosphere is also increased, which raises the recombination rate. Therefore, the over-population of the excited levels is lessened and the continuum begins to brighten. For cases with a stronger electron beam bombardment, the electron density in the heated atmosphere increases more and faster. This leads to an earlier brightening of the continuum, i.e., a shorter dimming. In the meantime, a higher electron beam flux can cause a greater population excess of the hydrogen excited levels, resulting in a stronger dimming. For cases with a quiet-Sun initial atmosphere, we can see that they show a longer dimming than those with a penumbral atmosphere. This is because as the flare evolves, the ratio of recombinations to photoionizations grows more slowly for the quiet-Sun cases, which present a relatively lower temperature in the upper chromosphere and thus a lower ratio of collisional rate to photoionization rate.

A threshold of detectable continuum enhancement has been used in the past to identify a WLF with the practical consideration of instrumental sensitivity in observations. Different from 3\% used by \citet{cheng2010radiative}, we assume a threshold of 2\% by taking into account the development of detectors in recent years. When the continuum contrast of a flare exceeds this value, this flare is regarded as a WLF. For the continuum at 3600\angstrom, an F10 electron beam can lead to a detectable enhancement for both two spectral indices in the quiet-Sun initial atmosphere. However, for the penumbral atmosphere, the smallest F9 in the flux parameter space is capable to reproduce a detectable enhancement. As regards the continuum at 4250\angstrom, when the quiet-Sun initial atmosphere is considered, only a 5F10 electron beam with spectral index $\delta=3$ can reproduce a required enhancement. While for the penumbral atmosphere, a 3F10 electron beam is required for the detectable continuum emission for both spectral indices. Note that the needed flux is somewhat greater than that of previously reported value from \citet{cheng2010radiative}, which is 3F9 for the continuum at 4300\angstrom\ for a disk center event. This discrepancy could be attributed to the difference in the heating function and heating duration. Here we can see that the flux threshold for continuum at 4250\angstrom\ is greater than that of 3600\angstrom. This is simply because that the continuum emission at 4250\angstrom\ (the Paschen continuum) originates from a deeper layer than the continuum at 3600\angstrom\ (the Balmer continuum) \citep{kowalski2015new}. 

As for the H$\alpha$ and Ly$\alpha$ lines, several previous studies have shown that their excess emission has a strong relationship with non-thermal electron beam bombardment \citep{allred2005radiative, rubiodacosta2011solar,kuridze2015h,hong2019response,prochazka2019hydrogen}. Our study broadly supports these studies and provides a theoretical basis to interpret future observations. It should be noted that a remarkable difference between the H$\alpha$ and Ly$\alpha$ lines lies in whether there is a dimming exhibited in their intensity curves. The H$\alpha$ dimming exists and originates from the increase of hydrogen population of the first excited state. When the electrons penetrate into the lower atmosphere, local hydrogen atoms will be excited to higher energy levels. At the beginning of the heating, the increase of the population of first excited state typically overweighs the higher energy states. Therefore, more hydrogen atoms in the first excited state increases the opacity and leads to more absorption at the H$\alpha$ line, which causes the dimming of the line \citep{abbett1999dynamic, allred2005radiative}. But for the Ly$\alpha$ line, which is generated by the transition between the first excited state and the ground state of the hydrogen atom, since the ground state is less populated when the atmosphere is bombarded by non-thermal electrons, it does not show any dimming when the heating begins.

We have also shown that the H$\alpha$ and Ly$\alpha$ lines can exhibit an asymmetry and even an asymmetry change when the flaring atmosphere is heated by non-thermal beam electrons. In particular, the Ly$\alpha$ line asymmetry changes in coincidence with the light curve ``platform". This intensity ``platform" has also been shown in previous simulations and named as a ``dip" \citep{hong2019response}, which is believed to be a mark of non-thermal electron beam heating. For the H$\alpha$ line, an asymmetry change can also be seen, say, in the model 5F10d3Q at about 10\,s. Note that the ``platform" in the H$\alpha$ light curves is generally inconspicuous, which may be due to the fact that the asymmetry change happens near the peak time of electron flux, which also flattens the H$\alpha$ light curve to some extent. These characteristics shown in the line profiles and also the light curves could be served as a reference for the future high-resolution observations.

Considering that the observational wavebands of the future LST/ASO-S telescope include the Ly$\alpha$ line and 3600\angstrom, our simulations can be served as a guidance as how the observations should be done. The high-cadence of LST is expected to observe the dimming or black-light flares at a time resolution of 4\,s (for 3600\angstrom, the time resolution can be as short as 1\,s), which have rarely been observed before. 
Moreover, if our simulations are combined with the observations of the future LST/ASO-S and also the ONSET telescope, which include comprehensive observations in the Ly$\alpha$, 3600\angstrom\ and 4250\angstrom, we could expect a diagnosis of the heating mechanism and energy transportation in solar WLFs. 

In should be also mentioned that in this work, we only focus on the beam heating in a single loop. However, the real flares are actually composed of many small-scale threads. Therefore, to facilitate the comparison between observations and numerical simulations, multithreaded simulations are expected. A multithreaded hydrodynamic model has been employed by \citet{reep2019efficient} to describe the heating in flares, and some observed spectral properties can be explained with such a model. By considering a series of successively heated loops in the model, \citet{reep2020simulating} reproduced the observed light curves, quasi-periodic pulsations, and soft X-ray spectra. We expect that some effects might be enhanced and others might be reduced if using a multithreaded model. It is worth trying that kind of model in the future to study solar WLFs. 

Finally, our main conclusions can be outlined as follows:

\begin{enumerate}

\item The non-thermal electron beam heating model can clearly show enhancements in the continua at 3600{\,\AA} and 4250{\,\AA} and in the H$\alpha$ and Ly$\alpha$ lines.   
\item A larger electron beam flux, a smaller spectral index or a penumbral initial atmosphere leads to a stronger emission increase at 3600{\,\AA}, 4250{\,\AA} and in the H$\alpha$ line, whereas for the Ly$\alpha$ line, a larger spectral index and a quiet-Sun initial atmosphere are preferable conditions for the line intensity increase. 
\item It is also notable that the continua at 3600\angstrom\ and 4250\angstrom\ and the H$\alpha$ line exhibit a dimming at the beginning of the heating and reach their peak emissions later than the peak time of the heating function. However, the Ly$\alpha$ line does not show a dimming and reaches its peak emission at nearly the same time as the heating function.
\item Both the H$\alpha$ and Ly$\alpha$ lines show a change from red asymmetry to blue asymmetry when an intense heating exists. Such a change occurs later for H$\alpha$ compared with Ly$\alpha$, and the asymmetry change in H$\alpha$ line profiles is not as significant as that in Ly$\alpha$ line profiles.

\end{enumerate}

\begin{acknowledgements}
This work was supported by NSFC under grants 11873095, 11903020, 11733003, and U1731241, and by the CAS Strategic Pioneer Program on Space Science under grants XDA15052200, XDA15320103, and XDA15320301. Y.L. is also supported by the CAS Pioneer Talents Program for Young Scientists.
\end{acknowledgements}


\label{lastpage}


\begin{thebibliography}{99}

\bibitem[Abbett \& Hawley(1999)]{abbett1999dynamic} Abbett, W.~P., \& Hawley, S.~L.\ 1999, \apj, 521, 906

\bibitem[Aboudarham \& Henoux(1986)]{aboudarham1986energy} Aboudarham, J., \& Henoux, J.~C.\ 1986, \aap, 156, 73

\bibitem[Allred et al.(2005)]{allred2005radiative} Allred, J.~C., Hawley, S.~L., Abbett, W.~P., et al.\ 2005, \apj, 630, 573

\bibitem[Allred et al.(2006)]{allred2006radiative} Allred, J.~C., Hawley, S.~L., Abbett, W.~P., et al.\ 2006, \apj, 644, 484

\bibitem[Allred et al.(2015)]{allred2015unified} Allred, J.~C., Kowalski, A.~F., \& Carlsson, M.\ 2015, \apj, 809, 104

\bibitem[Brown(1971)]{brown1971deduction} Brown, J.~C.\ 1971, \solphys, 18, 489

\bibitem[Carlsson \& Stein(1992)]{carlsson1992non} Carlsson, M., \& Stein, R.~F.\ 1992, \apjl, 397, L59

\bibitem[Carlsson \& Stein(1995)]{carlsson1995does} Carlsson, M., \& Stein, R.~F.\ 1995, \apjl, 440, L29

\bibitem[Carlsson \& Stein(1997)]{carlsson1997formation} Carlsson, M., \& Stein, R.~F.\ 1997, \apj, 481, 500

\bibitem[Carlsson \& Stein(2002)]{carlsson2002dynamic} Carlsson, M., \& Stein, R.~F.\ 2002, \apj, 572, 626

\bibitem[Carrington(1859)]{carrington1859description} Carrington, R.~C.\ 1859, \mnras, 20, 13

\bibitem[Chen \& Ding(2005)]{chen2005relationship} Chen, Q.~R., \& Ding, M.~D.\ 2005, \apj, 618, 537

\bibitem[Chen \& Ding(2006)]{chen2006footpoint} Chen, Q.~R., \& Ding, M.~D.\ 2006, \apj, 641, 1217

\bibitem[Cheng et al.(2010)]{cheng2010radiative} Cheng, J.~X., Ding, M.~D., \& Carlsson, M.\ 2010, \apj, 711, 185

\bibitem[Dere et al.(2019)]{dere2019chianti} Dere, K.~P., Del Zanna, G., Young, P.~R., et al.\ 2019, \apjs, 241, 22

\bibitem[Dere et al.(1997)]{dere1997chianti} Dere, K.~P., Landi, E., Mason, H.~E., et al.\ 1997, \aaps, 125, 149

\bibitem[Ding \& Fang(1989)]{ding1989semi-empirical} Ding, M.~D., \& Fang, C.\ 1989, \aap, 225, 204

\bibitem[Ding \& Fang(1996)]{ding1996possible} Ding, M.~D., \& Fang, C.\ 1996, \solphys, 166, 437

\bibitem[Ding et al.(1994)]{ding1994optical} Ding, M.~D., Fang, C., Gan, W.~Q., et al.\ 1994, \apj, 429, 890

\bibitem[Dorfi \& Drury(1987)]{dorfi1987simple} Dorfi, E.~A., \& Drury, L.~O.\ 1987, Journal of Computational Physics, 69, 175

\bibitem[Emslie \& Sturrock(1982)]{emslie1982temperature} Emslie, A.~G., \& Sturrock, P.~A.\ 1982, \solphys, 80, 99

\bibitem[Fang et al.(2013)]{fang2013new} Fang, C., Chen, P.-F., Li, Z., et al.\ 2013, Research in Astronomy and Astrophysics, 13, 1509-1517

\bibitem[Fletcher et al.(2007)]{fletcher2007trace} Fletcher, L., Hannah, I.~G., Hudson, H.~S., et al.\ 2007, \apj, 656, 1187

\bibitem[Fletcher \& Hudson(2008)]{fletcher2008impulsive} Fletcher, L., \& Hudson, H.~S.\ 2008, \apj, 675, 1645

\bibitem[Gan \& Mauas(1994)]{gan1994atmospheric} Gan, W.~Q., \& Mauas, P.~J.~D.\ 1994, \apj, 430, 891

\bibitem[Gustafsson(1973)]{gustafsson1973} Gustafsson, B.\ 1973, Uppsala Astron. Obs. Ann., 5, 1

\bibitem[Hao et al.(2017)]{hao2017circular} Hao, Q., Yang, K., Cheng, X., et al.\ 2017, Nature Communications, 8, 2202

\bibitem[Hong et al.(2017)]{hong2017radyn} Hong, J., Carlsson, M., \& Ding, M.~D.\ 2017, \apj, 845, 144

\bibitem[Hong et al.(2018)]{hong2018non} Hong, J., Ding, M.~D., Li, Y., et al.\ 2018, \apjl, 857, L2

\bibitem[Hong et al.(2019)]{hong2019response} Hong, J., Li, Y., Ding, M.~D., et al.\ 2019, \apj, 879, 128

\bibitem[Hudson(1972)]{hudson1972thick} Hudson, H.~S.\ 1972, \solphys, 24, 414

\bibitem[Hudson(2016)]{hudson2016chasing} Hudson, H.~S.\ 2016, \solphys, 291, 1273

\bibitem[Hudson et al.(1992)]{hudson1992white} Hudson, H.~S., Acton, L.~W., Hirayama, T., et al.\ 1992, \pasj, 44, L77

\bibitem[Hudson et al.(2006)]{hudson2006white} Hudson, H.~S., Wolfson, C.~J., \& Metcalf, T.~R.\ 2006, \solphys, 234, 79

\bibitem[Isobe et al.(2007)]{isobe2007flare} Isobe, H., Kubo, M., Minoshima, T., et al.\ 2007, \pasj, 59, S807

\bibitem[Jess et al.(2008)]{jess2008do} Jess, D.~B., Mathioudakis, M., Crockett, P.~J., et al.\ 2008, \apjl, 688, L119

\bibitem[Jing et al.(2008)]{jing2008spatial} Jing, J., Chae, J., \& Wang, H.\ 2008, \apjl, 672, L73

\bibitem[Kane et al.(1985)]{kane1985characteristics} Kane, S.~R., Love, J.~J., Neidig, D.~F., et al.\ 1985, \apjl, 290, L45

\bibitem[Kerr et al.(2019a)]{kerr2019a} Kerr, G.~S., Allred, J.~C., \& Carlsson, M.\ 2019, \apj, 883, 57

\bibitem[Kerr et al.(2019b)]{kerr2019b} Kerr, G.~S., Carlsson, M., \& Allred, J.~C.\ 2019, \apj, 885, 119

\bibitem[Kowalski et al.(2015)]{kowalski2015new} Kowalski, A.~F., Hawley, S.~L., Carlsson, M., et al.\ 2015, \solphys, 290, 3487

\bibitem[Kowalski et al.(2017)]{kowalski2017hydrogen} Kowalski, A.~F., Allred, J.~C., Uitenbroek, H., et al.\ 2017, \apj, 837, 125

\bibitem[Krucker et al.(2015)]{krucker2015co} Krucker, S., Saint-Hilaire, P., Hudson, H.~S., et al.\ 2015, \apj, 802, 19

\bibitem[Kuhar et al.(2016)]{kuhar2016correlation} Kuhar, M., Krucker, S., Mart{\'\i}nez Oliveros, J.~C., et al.\ 2016, \apj, 816, 6

\bibitem[Kuridze et al.(2015)]{kuridze2015h} Kuridze, D., Mathioudakis, M., Sim{\~o}es, P.~J.~A., et al.\ 2015, \apj, 813, 125

\bibitem[Lee et al.(2017)]{lee2017irisa} Lee, K.-S., Imada, S., Watanabe, K., et al.\ 2017, \apj, 836, 150

\bibitem[Li et al.(2019)]{li2019lyman} Li, H., Chen, B., Feng, L., et al.\ 2019, Research in Astronomy and Astrophysics, 19, 158

\bibitem[Li et al.(1997)]{li1997magnetic} Li, X.~Q., Song, M.~T., Hu, F.~M., et al.\ 1997, \aap, 320, 300

\bibitem[Liu et al.(2001)]{liu2001enhanced} Liu, Y., Ding, M.~D., \& Fang, C.\ 2001, \apjl, 563, L169

\bibitem[Machado \& Rust(1974)]{machado1974analysis} Machado, M.~E., \& Rust, D.~M.\ 1974, \solphys, 38, 499

\bibitem[Machado et al.(1978)]{machado1978} Machado, M.~E., Emslie, A.~G., \& Brown, J.~C.\ 1978, \solphys, 58, 363

\bibitem[Machado et al.(1986)]{machado1986mechanism} Machado, M.~E., Emslie, A.~G., \& Mauas, P.~J.\ 1986, \aap, 159, 33

\bibitem[Machado et al.(1989)]{machado1989radiative} Machado, M.~E., Emslie, A.~G., \& Avrett, E.~H.\ 1989, \solphys, 124, 303

\bibitem[Matthews et al.(2003)]{matthews2003catalogue} Matthews, S.~A., van Driel-Gesztelyi, L., Hudson, H.~S., et al.\ 2003, \aap, 409, 1107

\bibitem[McTiernan \& Petrosian(1990)]{mctiernan1990behavior} McTiernan, J.~M., \& Petrosian, V.\ 1990, \apj, 359, 524

\bibitem[Metcalf et al.(2003)]{metcalf2003trace} Metcalf, T.~R., Alexander, D., Hudson, H.~S., et al.\ 2003, \apj, 595, 483

\bibitem[Namekata et al.(2017)]{namekata2017statisticala} Namekata, K., Sakaue, T., Watanabe, K., et al.\ 2017, \apj, 851, 91

\bibitem[Neidig et al.(1993)]{neidig1993solar} Neidig, D.~F., Kiplinger, A.~L., Cohl, H.~S., et al.\ 1993, \apj, 406, 306

\bibitem[Neidig \& Kane(1993)]{neidig1993energetics} Neidig, D.~F., \& Kane, S.~R.\ 1993, \solphys, 143, 201

\bibitem[Pereira \& Uitenbroek(2015)]{pereira2015rh} Pereira, T.~M.~D., \& Uitenbroek, H.\ 2015, \aap, 574, A3

\bibitem[Proch{\'a}zka et al.(2018)]{prochazka2018reproducing} Proch{\'a}zka, O., Reid, A., Milligan, R.~O., et al.\ 2018, \apj, 862, 76

\bibitem[Proch{\'a}zka et al.(2019)]{prochazka2019hydrogen} Proch{\'a}zka, O., Reid, A., \& Mathioudakis, M.\ 2019, \apj, 882, 97

\bibitem[Reep et al.(2019)]{reep2019efficient} Reep, J.~W., Bradshaw, S.~J., Crump, N.~A., et al.\ 2019, \apj, 871, 18

\bibitem[Reep et al.(2020)]{reep2020simulating} Reep, J.~W., Warren, H.~P., Moore, C.~S., et al.\ 2020, \apj, 895, 30

\bibitem[Rubio da Costa(2011)]{rubiodacosta2011solar} Rubio da Costa, F.\ 2011, Ph.D. Thesis

\bibitem[Rust \& Hegwer(1975)]{rust1975analysis} Rust, D.~M., \& Hegwer, F.\ 1975, \solphys, 40, 141

\bibitem[Song et al.(2018)]{song2018observations} Song, Y.~L., Guo, Y., Tian, H., et al.\ 2018, \apj, 854, 64

\bibitem[Sun et al.(2013)]{sun2013hot} Sun, X., Hoeksema, J.~T., Liu, Y., et al.\ 2013, \apj, 778, 139

\bibitem[Tsuneta et al.(1991)]{tsuneta1991soft} Tsuneta, S., Acton, L., Bruner, M., et al.\ 1991, \solphys, 136, 37

\bibitem[Uitenbroek(2001)]{uitenbroek2001multilevel} Uitenbroek, H.\ 2001, \apj, 557, 389

\bibitem[Uitenbroek(2002)]{uitenbroek2002effect} Uitenbroek, H.\ 2002, \apj, 565, 1312

\bibitem[Ulmschneider et al.(1987)]{ulmschneider1987acoustic} Ulmschneider, P., Muchmore, D., \& Kalkofen, W.\ 1987, \aap, 177, 292

\bibitem[Vernazza et al.(1981)]{vernazza1981structure} Vernazza, J.~E., Avrett, E.~H., \& Loeser, R.\ 1981, \apjs, 45, 635

\bibitem[Wang(2009)]{wang2009study} Wang, H.-M.\ 2009, Research in Astronomy and Astrophysics, 9, 127

\bibitem[Watanabe et al.(2017)]{watanabe2017characteristics} Watanabe, K., Kitagawa, J., \& Masuda, S.\ 2017, \apj, 850, 204

\bibitem[Watanabe et al.(2010)]{watanabe2010g} Watanabe, K., Krucker, S., Hudson, H., et al.\ 2010, \apj, 715, 651

\bibitem[Watanabe \& Imada(2020)]{watanabe2020white} Watanabe, K., \& Imada, S.\ 2020, \apj, 891, 88

\bibitem[Xu et al.(2006)]{xu2006high} Xu, Y., Cao, W., Liu, C., et al.\ 2006, \apj, 641, 1210

\end{thebibliography}
\end{document}